\documentclass[letterpaper, 10 pt, conference]{ieeeconf}  

\IEEEoverridecommandlockouts                              
\usepackage[utf8]{inputenc}
\usepackage[T1]{fontenc}
\usepackage{graphicx}
\usepackage{array}
\usepackage{tabularx}
\usepackage{caption}
\usepackage{subcaption}
\usepackage{graphics}

\title{\LARGE \bf
Personality Trait Recognition using ECG Spectrograms and Deep Learning}

\author{Muhammad Mohsin Altaf$^{1}$,  Saadat Ullah Khan $^{1}$, Muhammad Majid $^{1}$, and Syed Muhammad Anwar$^{2}$
\thanks{*This work was not supported by any organization}
\thanks{$^{1}$Muhammad Mohsin Altaf, Saadat Ullah Khan, and Muhammad Majid are with Department of Computer Engineering, University of Engineering and Technology, Taxila, Pakistan.}%
\thanks{$^{2}$Syed Muhammad Anwar is with with Sheikh Zayed Institute for Pediatric Surgical Innovation, Children’s National Hospital, Washington, DC and School of Medicine and Health Sciences, George Washington University, Washington, DC.}%
}

\begin{document}

\maketitle

\begin{abstract}
This paper presents an innovative approach to recognizing personality traits using deep learning (DL) methods applied to electrocardiogram (ECG) signals. Within the framework of detecting the big five personality traits model encompassing extra-version, neuroticism, agreeableness, conscientiousness, and openness, the research explores the potential of ECG-derived spectrograms as informative features. Optimal window sizes for spectrogram generation are determined, and a convolutional neural network (CNN), specifically Resnet-18, and visual transformer (ViT) are employed for feature extraction and personality trait classification. The study utilizes the publicly available ASCERTAIN dataset, which comprises various physiological signals, including ECG recordings, collected from 58 participants during the presentation of video stimuli categorized by valence and arousal levels. The outcomes of this study demonstrate noteworthy performance in personality trait classification, consistently achieving F1-scores exceeding 0.9 across different window sizes and personality traits. These results emphasize the viability of ECG signal spectrograms as a valuable modality for personality trait recognition, with Resnet-18 exhibiting remarkable effectiveness in discerning distinct personality traits.
\end{abstract}

\begin{keywords}
Personality Recognition, Electrocardiogram, Spectrogram, Deep Learning.
\end{keywords}

\section{INTRODUCTION}
\begin{figure*}[t]
    \centering
    \includegraphics[width=0.83\textwidth]{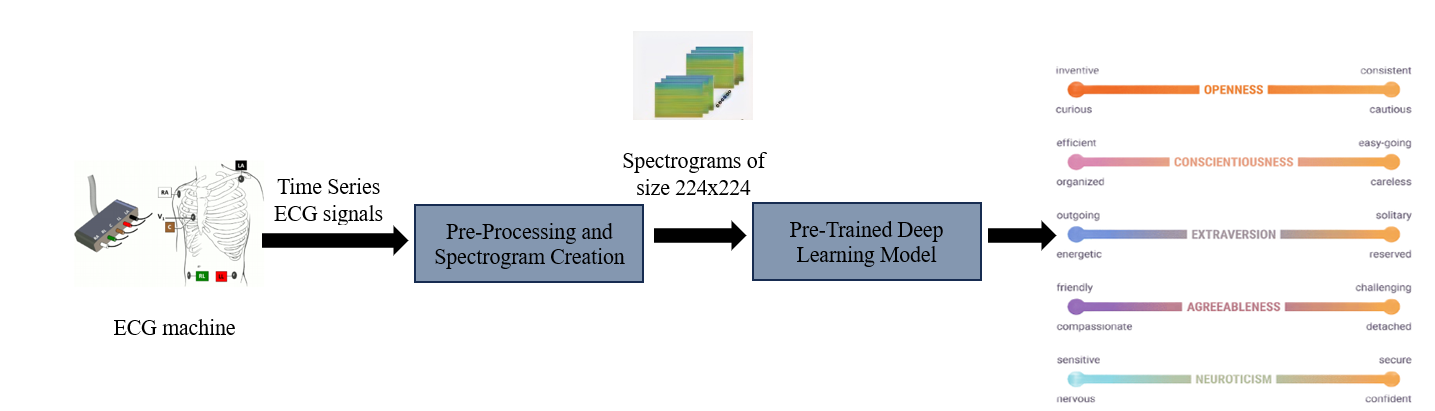}
    \caption{Our proposed approach for personality recognition using ECG spectrograms and deep learning models.}
    \label{fig:1}  
\end{figure*}

The study of human personality is fundamental to understanding behavior, cognition, and social interactions. Recognizing and categorizing personality traits has implications across a spectrum of fields, from psychology and healthcare to personalized user experiences in technology. Accurate personality recognition offers the potential to tailor interventions, services, and interactions to individual preferences, thereby enhancing well-being and optimizing outcomes \cite{willroth2023personality}. The study of personality encompasses a multifaceted landscape, with numerous models and theories developed over decades. One such widely recognized framework is the "Big Five" personality traits model \cite{shui2023personality}, which encompasses extra-version (energetic vs. reserved), neuroticism (nervous vs. confident), agreeableness (compassionate vs. detached), conscientiousness (organized vs. careless), and openness (curious/creative vs. cautious/conservative). These traits capture essential dimensions of human personality, providing a basis for understanding individual differences and behaviors \cite{simha2020big}.

Traditionally, an assessment for personality traits has relied heavily on self-reported questionnaires and behavioral observations. While these methods have proven valuable, they also introduce subjectivity and potential biases \cite{stachl2021computational}. Advances in technology and data-driven methodologies have opened doors to novel approaches to personality recognition, offering the potential for more objective and data-rich assessments. In the pursuit of accurate and reliable personality recognition, researchers have explored a spectrum of methods. These range from classical psychological assessments and observational techniques to cutting-edge machine learning algorithms \cite{smith2019moving}. Yet, challenges persist in achieving consistently high accuracy, particularly when dealing with nuanced and context-dependent personality dimensions \cite{khan2023exploring}.

Recent years have witnessed the convergence of physiological signal analysis and machine learning, offering exciting prospects for personality recognition \cite{kaklauskas2022review}. One such physiological signal, the electrocardiogram (ECG)- which measures the electrical activity of the heart- has garnered attention for its potential to reveal hidden insights into personality traits \cite{lin2023review}. ECG signals are not only easily accessible but also inherently linked to emotional states, making them a promising modality for personality recognition \cite{panahi2021application}.

This research harnesses the power of ECG signals, image processing, and deep learning techniques to advance the field of personality recognition. The data utilized in this study is taken from a public repository: ASCERTAIN \cite{7736040}, which comprises an array of physiological signals, including ECG recordings, obtained from 58 participants exposed to a variety of video stimuli categorized by arousal and valance levels. The intricate interplay between physiological signals and emotional responses underpins our exploration of personality traits. Our methodology involves the generation of spectrograms from ECG signals using carefully optimized window sizes. Convolutional Neural Networks (CNNs), specifically the Resnet-18 architecture \cite{he2016deep} and visual transformer architecture \cite{dosovitskiy2020image}, serve as the backbone for feature extraction and personality trait classification. In particular, our major contribution is utilizing spectrograms from physiological signals (ECG) and train large deep learning models to accurately classify personality traits.

The rest of the paper is organized as follows. Section II describes the proposed methodology. Experimental results are presented in Section III followed by a conclusion in Section IV.


\section{PROPOSED METHODOLOGY}
\label{sec:format}
A comprehensive block diagram of our proposed methodology is shown in Fig. 1, which uses ECG signal analysis and deep learning techniques to achieve precise personality recognition. Commencing with the robust ASCERTAIN dataset \cite{7736040}, which encompasses a spectrum of physiological signals, prominently featuring ECG recordings from participants exposed to video stimuli of varying valence and arousal levels. The core of our methodology resides in the transformation of ECG signals into spectrograms, a process meticulously optimized for feature-rich representation. The subsequent step involves the deployment of CNNs, with a particular focus on the Resnet-18 \cite{he2016deep} and ViT architecture's capacity to interpret complex patterns from spectrogram data. The seamless integration of these components enables us to establish a robust model capable of discerning personality traits from ECG signals. The details of each block are as follows.

\subsection{Electrocardiogram Data}
\label{sec:majhead}

The ASCERTAIN dataset comprises a diverse range of physiological signals, including ECG recordings, collected from 58 participants exposed to video stimuli (36 videos) categorized in different categories based on valence and arousal levels. In particular, there are four subcategories of these 36 video clips. Clip 1 to 9 is categorized into High Arousal and High Valance (HAHV), Clip 10 to 18 Low Arousal and High Valance (LAHV), Clips 19 to 27 Low Arousal and Low Valance (LALV), and 28 to 36 High Arousal and Low Valance (HALV) clips \cite{7736040}. ECG signals from the right and left arm were recorded at a sampling rate of 256Hz. In this dataset, ECG data captures real-time physiological responses to stimuli, offering a unique opportunity to investigate the connection between emotional reactions, physiological signals, and personality traits \cite{stachl2021computational}. All the subjects were asked to fill out a questionnaire with 50 questions and based on replies gathered the personality traits of all subjects were determined. The mean scores for each personality feature were calculated to group people into personality types. Introverts were defined as participants with mean personality trait scores lower or equal to  4.31, while extroverts were defined as participants with mean personality trait scores higher than 4.31. Similarly, Non-agreeables were defined as participants with mean personality trait scores lower or equal to  5.09, while agreeables were defined as participants with mean personality trait scores higher than 5.09. Non-conscientiousness was defined as participants with mean personality trait scores lower or equal to  5.14, while conscientiousness was defined as participants with mean personality trait score higher than 5.14. Emotionally unstable was defined as participants with mean personality trait scores lower or equal to  4.14, while emotionally stable was defined as participants with mean personality trait scores higher than 4.14. Open-minded was defined as participants with mean personality trait scores lower or equal to 4.95, while close-minded was defined as participants with mean personality trait scores higher than 4.95 \cite {7736040}.

\begin{figure}[t]
    \centering
    \includegraphics[width = 0.45\textwidth]{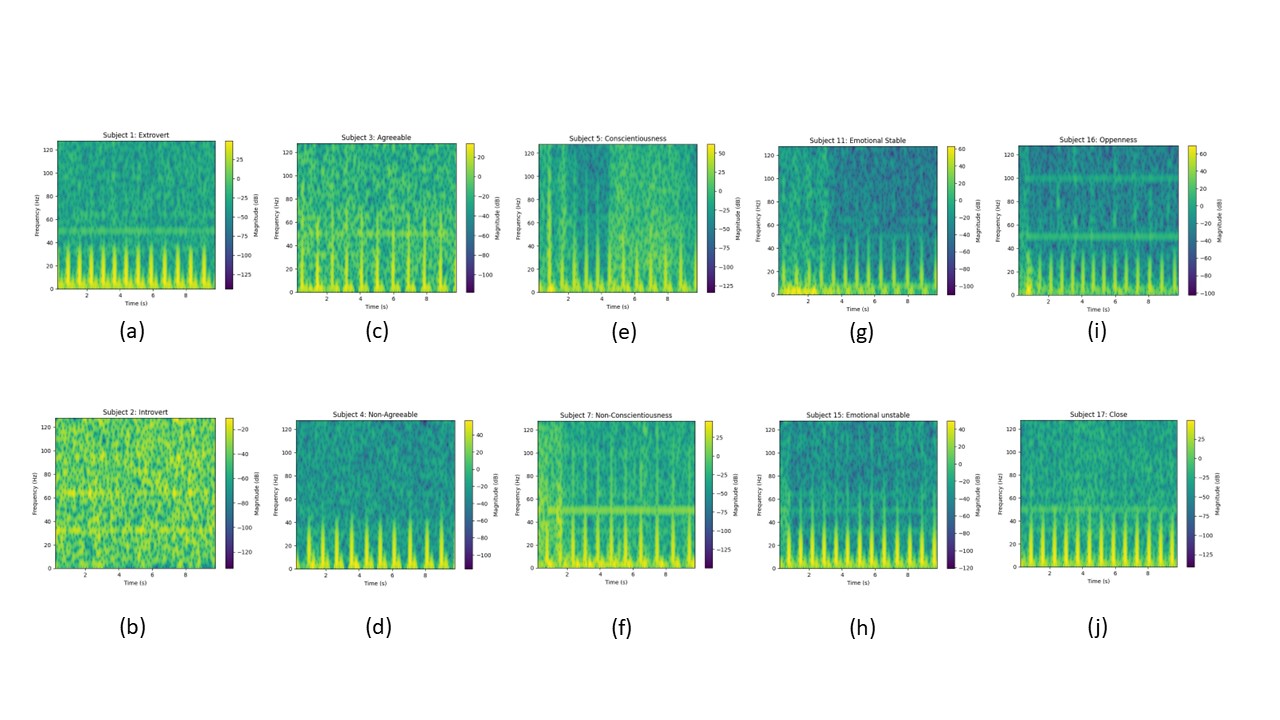}
    \caption{Spectrograms of personality traits (a) Extrovert (b) Introvert (c) Agreeable (d) Non-Agreeable (e) Conscientiousness (f) Non-conscientiousness (g) Emotionally Stable (h) Emotionally Unstable (i) Open-minded (j) Close-minded. }
    \label{fig:2}
\end{figure}

\begin{table*}[t]
\centering
\caption{Performance of the proposed personality recognition framework using ECG spectrograms and deep learning models.}
\begin{tabular}{|c|*{7}{c|}}
\hline
 & & \multicolumn{3}{c|}{Window size = 100, overlap = $89\%$ } & \multicolumn{3}{c|}{Window size = 327, overlap = $98\%$ }\\
\cline{3-8}
Deep Learning Models& Personality Traits & Precision & Recall & F1-Score & Precision & Recall & F1-Score \\
\cline{1-8}
     &Extraversion&0.92 &0.95 &0.94 &0.92&0.96&0.94\\
\cline{2-8}
     & Agreeableness&0.95&0.89&0.92&0.93&0.91&0.92\\
\cline{2-8}
    Resnet-18 \cite{he2016deep}&Conscientiousness&0.91&0.97&0.94&0.92&0.93&0.92\\
\cline{2-8}
     &Emotional Stability&0.96&0.90&0.93&0.96&0.90&0.93\\
\cline{2-8}
     &Openness&0.92&0.95&0.94&0.92&0.95&0.93\\
\cline{2-8}
\hline
\cline{2-8}
 &Extraversion&0.82&0.76&0.81&0.85&0.88&0.86\\
\cline{2-8}
 & Agreeableness&0.78&0.78&0.76&0.86&0.84&0.84\\
\cline{2-8}
ViT \cite{dosovitskiy2020image}&Conscientiousness&0.84&0.80&0.81&0.88&0.88&0.87\\
\cline{2-8}
 &Emotional Stability&0.84&0.85&0.86&0.93&0.82&0.87\\
\cline{2-8}
 &Openness&0.80&0.83&0.81&0.84&0.88&0.86\\
\hline
\end{tabular}
\end{table*}

\subsection{Pre-Processing and Spectrogram Generation}
\label{sec:majhead}
We divided the ECG recordings of each subject from the ASCERTAIN dataset into various segments each consisting of 10 seconds of the ECG signal. All these segments are then used for the computation of the spectrogram. The Short-Time Fourier Transform (STFT) can be used to determine the frequency and phase properties of particular signal segments and monitor how they change over time. The STFT includes applying several Fourier transformations to signal segments, each of which is then subjected to structural windowing. This connection and the STFT's underlying mathematical model can be represented in Eq. 1 as,
\begin{equation}
X_m(\omega)=\sum_{n=-\infty}^{\infty}{x\left(n+mR\right)w(n)e^{-\omega(n+mR)}}    
\end{equation}
where $x$ is the input signal at time $n$, $w(n)$ is the window of length $m$, $R$ is the size of hop between the successive DTFTs, and $X_m(\omega)$ is the DTFT of the windowed data \cite{khan2023motor}.

In this work, we used the spectrogram of a fixed size of $224 \times 224$ using different window sizes and overlapping samples in the window. To fix the height of the spectrogram at $224$ we set the number of discrete Fourier transform points at $447$ and subsequently used the window and overlap length to set the width of the spectrogram at $224$. The size of the spectrogram is thus congruent with the input dimensions of the deep learning models used, which is fixed at $224 \times 224$. Blackman window is used as the windowing function to calculate the spectrograms and we took a natural log of the magnitude square of each spectrogram. Fig. 2 shows the spectrograms of binary classified personality traits of different subjects for different personality traits. It can be observed that different personality traits result in visually different spectrograms that can be used to extract features for better classification of the underlying personality trait.


\subsection{Pre-Trained Deep Learning Models}
\label{sssec:subsubhead}

We employ state-of-the-art pre-trained architectures to extract intricate features from spectrogram data derived from ECG signals. Specifically, our exploration encompasses the Resnet-18 \cite{he2016deep} and  Vision Transformer (ViT) models\cite{dosovitskiy2020image}, each renowned for its unique set of attributes. Resnet-18, a proven workhorse in image data analysis, excels in capturing intricate spatial features, while the Vision Transformer offers a transformative approach with its attention mechanisms, allowing for comprehensive global context analysis. These models, initially trained on extensive ECG signals spectrograms of the ASCERTAIN dataset come endowed with an extensive knowledge base on binary classification of the personality traits. Following this pre-training, fine-tuning our spectrogram data further enhances model adaptation, resulting in optimal performance.

\section{EXPERIMENTAL RESULTS AND DISCUSSION}

A publically available ASCERTAIN dataset was used for the binary classification of five personality traits. These five personality traits are extraversion (Ext), agreeableness (Agr), openness (Ope), conscientiousness (Con), and emotionally stable (EmS). We chose the ECG modality from the ASCERTAIN dataset, which is collected from the left arm for the calculation of spectrograms. The computation of the spectrograms and the training of the deep learning models were performed on a personal computer with an Intel®Xeon®W-2265 CPU running at 3.50GHz, 64 GB of physical memory, and a Nvidia®RTX TM A5000 GPU.

The selection of the window length and the number of overlapping samples are two important parameters to take into account while calculating the spectrograms. There are many combinations of the window size and overlap window through which we can get the spectrogram of the dimensions $224 \time 224$. In this work, we have used two different window sizes i.e., $100$ and $327$ with overlapping of 89\% (89 samples) and 96\% (321 samples) respectively. We replicated the values of the spectrogram across all the input channels of the deep learning models. We chose Resnet-18, and Vision Transformer for the binary classification of five personality traits. 10-fold cross-validation was used for the fine-tuning of these pre-trained models. We used 3 epochs with 16 as batch size. Cross entropy loss and stochastic gradient descent (SGD) were chosen as the loss function and optimizer for fine-tuning, respectively. To minimize overfitting and underfitting of the models we used 0.9 as momentum and weight decay as 0.001. We used 0.001 as the learning rate. Batch normalization was used to improve the performance of the models. For the binary classification of the personality traits, we replaced the last layers of the pre-trained models with a linear layer consisting of two neurons.

Table 1 presents the precision, recall, and F1-score for different personality traits, and it compares the performance of two deep learning models, ResNet-18 and ViT (Vision Transformer), across these traits for both window sizes i.e., window size 100 with $89\%$ overlapping, and window size 327 with $98\%$ overlapping. It can be observed from the results that ResNet-18 has demonstrated strong performance in personality recognition across various traits. Particularly its high precision and recall values, along with consistent F1 scores greater than 0.9 for all personality traits. These results suggest that ResNet-18 effectively captures and classifies personality traits based on ECG spectrograms, showcasing its strong performance in the personality recognition framework. For ViT model, we argue that a larger data size would be needed  for better training.

\begin{table}
\centering
\caption{Performance comparison of the proposed method using ECG spectrograms constructed from window size of 327 with $98\%$ overlap and Resnet-18 deep learning model with other methods in terms of F1-score for all, HAHV, LAHV, LAHV, HALV videos.}
\vspace{4mm}
\begin{tabular}{|c|c|c|c|c|c|}
\hline

& Personality  & Proposed &  & SVM & SVM \\
Video&  Traits & Resnet-18 & NB & (Lin) & (RBF) \\

&  &  & \cite{7736040} & \cite{7736040} & \cite{7736040} \\

\cline{1-6}
     &Ext&0.94&0.56&0.06&0.53\\
\cline{2-6}
     & Agr&0.92&0.55&0.45&0.32\\
\cline{2-6}
    ALL&Con&0.92&0.60&0.51&0.55\\
\cline{2-6}
     &EmS&0.93&0.53&0.60&0.58\\
\cline{2-6}
     &Ope&0.93&0.48&0.35&0.49\\
\cline{2-6}

\hline
\cline{2-6}
 &Ext&0.90&0.59&0.00&0.56\\
\cline{2-6}
 & Agr&0.90&0.48&0.29&0.55\\
\cline{2-6}
HAHV&Con&0.89&0.50&0.32&0.52\\
\cline{2-6}
 &EmS&0.90&0.55&0.46&0.60\\
\cline{2-6}
 &Ope&0.89&0.45&0.34&0.55\\
\hline
\cline{1-6}
     &Ext&0.89&0.55&0.02&0.53\\
\cline{2-6}
     & Agr&0.86&0.58&0.45&0.60\\
\cline{2-6}
    LAHV&Con&0.89&0.70&0.78&0.74\\
\cline{2-6}
     &EmS&0.87&0.55&0.46&0.41\\
\cline{2-6}
     &Ope&0.89&0.56&0.42&0.49\\
\cline{2-6}

\hline
\cline{2-6}
 &Ext&0.91&0.58&0.10&0.49\\
\cline{2-6}
 & Agr&0.89&0.43&0.29&0.36\\
\cline{2-6}
LALV&Con&0.90&0.55&0.55&0.74\\
\cline{2-6}
 &EmS&0.89&0.53&0.58&0.50\\
\cline{2-6}
 &Ope&0.89&0.55&0.36&0.43\\
 \hline
\cline{2-6}
 &Ext&0.90&0.50&0.00&0.51\\
\cline{2-6}
 & Agr&0.89&0.51&0.32&0.62\\
\cline{2-6}
HALV&Con&0.91&0.57&0.57&0.62\\
\cline{2-6}
 &EmS&0.90&0.59&0.56&0.66\\
\cline{2-6}
 &Ope&0.89&0.45&0.33&0.50\\
\hline
\end{tabular}
\end{table}

Table 2 represents the performance comparison in terms of F1-scores of our proposed model using the Resnet-18 model for best-performing window size i.e., 327 with $98\%$ overlap, and three different models namely the Naive Bays (NB), Support Vector Machine (SVM) with linear, and SVM with Radial Basis Function (RBF) kernels performed alternatively by \cite{7736040}. The comparison is performed concerning all videos and for HAHV, LAHV, HALV, and LALV videos with an emphasis on the classification of the five core personality traits. Higher numbers indicate better model performance. When compared to the competing models, our personality recognition framework using ECG spectrograms consistently performs better. Particularly, our model outperformed the closest rival by a wide margin (0.56), achieving an accuracy of 0.94 for the extraversion trait. Similar trends are seen for agreeableness (0.92 versus 0.55), conscientiousness (0.92 versus 0.60), emotional stability (0.93 versus 0.53), and openness (0.93 versus 0.48).

\section{CONCLUSION}

This paper investigates the potential of analysis of ECG signals spectrograms for personality trait recognition, employing pre-trained deep learning models on a publicly available dataset. Notably, our research emphasizes the supremacy of deep learning models when applied to spectrograms derived from ECG signals, outperforming conventional methods from previous literature that necessitate feature extraction or employ raw ECG data as input. The results of this study demonstrate remarkable performance in personality trait recongition, consistently achieving F1-scores exceeding 0.9 across different window sizes and personality traits. Notably, ResNet-18 emerges as a highly effective model in discerning distinct personality traits based on ECG spectrogram data. In future, we intend to use self-sueprvised learning for training ViT models to furhter improve the performance.

\addtolength{\textheight}{-12cm}   




\end{document}